 \documentclass[preprint,NumberedRefs]{JASAnew}

\begin{document}

\title[]{Time-resolved measurement of acoustic density fluctuations using a phase-shifting Mach-Zehnder interferometer}
\author{Eita Shoji}
\email{ eita.shoji@tohoku.ac.jp}
\affiliation{Department of Mechanical Systems Engineering, Tohoku University, 6-6, Aramaki, Aoba-ku, Sendai, 980-8579, Japan.}

\author{Anis Maddi}
\email{anis.maddi@univ-lemans.fr}
\affiliation{Laboratoire d’Acoustique de l’Université du Mans (LAUM), UMR 6613,  Institut d’Acoustique - Graduate School (IA-GS), CNRS, Le Mans Université, France.}
 
\author{Guillaume Penelet}			
\affiliation{Laboratoire d’Acoustique de l’Université du Mans (LAUM), UMR 6613,  Institut d’Acoustique - Graduate School (IA-GS), CNRS, Le Mans Université, France.}

\author{Tetsushi Biwa}			
\affiliation{Department of Mechanical Systems Engineering, Tohoku University, 6-6, Aramaki, Aoba-ku, Sendai, 980-8579, Japan.}


\date{\today} 

\begin{abstract}
Phase-shifting interferometry is one of the optical measurement techniques that improves accuracy and resolution by incorporating a controlled phase shift into conventional optical interferometry. 
In this study, a four-step phase-shifting interferometer is developed to measure the spatio-temporal distribution of acoustic density oscillations of the gas next to a rigid plate. The experimental apparatus consists of a polarizing Mach-Zehnder interferometer with a polarization camera capable of capturing four polarization directions in one shot image and it is used to measure the magnitude and the phase of density fluctuations through a duct of rectangular cross-section connected to a loudspeaker. The results are compared with the well-established thermoacoustic theory describing the thermal coupling between acoustic oscillations and rigid boundaries, and the results show a very good agreement for various ratios of the (frequency-dependent) thermal boundary layer thickness to the plate spacing. This measurement technique could be advantageously employed to analyze more complex heat transfer processes involving the coupling of acoustic oscillations with rigid boundaries. 
\end{abstract}


\maketitle

\section{\label{sec:1} Introduction}

Acoustic wave propagation is generally studied in terms of pressure and velocity fluctuations, and several experimental techniques have been developed and successfully employed to access these physical quantities. This includes different systems like Laser Doppler Velocimetry (LDV) \cite{valiere2009development,gazengel2005measurement,valiere2000acoustic}, Particle Image Velocimetry (PIV) \cite{hireche2020experimental,blanc2003experimental,berson2008measurement,hann1997measurement} or Hot Wire Anemometry \cite{comte1976hot,jerbi2013acoustic} for the measurement of velocity, and different kinds of microphones for the pressure. However, acoustic waves are also characterized by fluctuations in density and temperature, which are \replaced{worth}{worthy} of investigation when considering the study of the thermal interaction between a wave and solid boundary. This interaction, for instance, is the fundamental process involved in thermoacoustic prime-movers and refrigerators, which \replaced{are developed since}{have been developed for} a few decades \cite{swift2003thermoacoustics,jin2015thermoacoustic}, but still need further investigation from a more fundamental standpoint. The linear thermoacoustic theory used for the design of such engines indeed fails in describing some complicated heat transport phenomena, notably those occurring at the ends of the heat exchangers \cite{gusev2001thermal,boluriaan2003acoustic,scalo2015linear} which give rise to a local distortion of temperature oscillations. More generally, as far as the heat exchange between a fluid submitted to acoustic oscillations and a solid boundary needs to be studied carefully, \replaced{it is useful to have at disposal a full-field}{it would be useful to have a full-field} and time-resolved measurement technique giving access to the spatial of distribution of the oscillating density or temperature.

To date, a handful of seminal works have been proposed to measure density and temperature fluctuations associated with acoustic wave propagation. More than two decades ago, Huelsz et al. proposed the use of cold-wire anemometry to measure the local temperature oscillations \cite{huelsz1998temperature} within the oscillating thermal boundary layers produced by an acoustic wave next to a wall. In the same period,  Wetzel et al. \cite{wetzel1999experimental,wetzel2000experimental,wetzel1998limitations} used another technique based on holographic interferometry combined with high-speed cinematography to measure the acoustically induced temperature oscillations between two parallel plates \replaced{submitted to}{subjected to} an axial temperature gradient. Later on, another optical technique based on using both PIV and PLIF (Planar Laser Induced Fluorescence) was successfully employed by Shi et al. \cite{shi2010application} to measure both the oscillating velocity and the oscillating temperature at the transition between a hot and a cold stack of plates. More recently, other experimental setups based on optical interferometry were employed to measure density and/or temperature fluctuations caused by acoustic wave propagation, in order to characterize the local nonlinear heat transport next to the heated side of the stack in a standing wave thermoacoustic prime-mover\cite{gong2018experimental,penelet2016measurement}, or to measure the spherically diverging shock waves emitted by a spark source\cite{yuldashev2015mach}.



Several of the techniques mentioned above are based on optical interferometry, which is an accurate full-field measurement technique exhibiting a high resolution. Furthermore, unlike probe measurements such as thermocouple or cold-wire sensors, it is a non-invasive and non-delayed measurement technique, which does not require a sophisticated calibration procedure. It has been successfully applied to a wide range of physical configurations \cite{hariharan2003optical}, ranging from fluid mechanics \cite{sugawara2020three,lauterborn1984modern,desse2017digital} to astronomy \cite{thompson2017interferometry,monnier2003optical}. In this respect, the Mach-Zehnder interferometer is considered one of the most common types of interferometer and has been in use for several decades. This interferometer relies on fringe pattern analysis to extract the phase difference generated between a reference beam and a test beam, the latter passing through the physical system to be measured. The phase extraction process generally involves a high level of processing and filtering, which can be avoided by using phase-shifting algorithms \cite{wyant1985recent,malacara2007optical}. 
Compared with conventional interferometers, phase-shifting interferometers offer several advantages: improvement of measurement resolution and noise reduction. In addition, it allows us to identify the wavefront \replaced{convexoconcave}{convexo-concave} of the test beam, i.e., the direction of density change, which cannot be determined by conventional optical interference fringes \cite{malacara2007optical}.

The concept of phase-shifting interferometry is the numerical computation of several acquired interferograms by introducing a phase shift between the reference and test beams. Although there are various methods of introducing the phase shift  \cite{malacara2018interferogram,shoji2015quantitative,shoji2012development,kanda2017measurement,torres2012development,shoji2015high}, the possibility of using digital cameras with several polarizers of different polarization directions placed in front of each pixel have become available \cite{ishikawa2016high,zhang2022preliminary,yang2022quantitative}. The use of polarization cameras eliminates the need for temporal synchronization between cameras which were required previously by some existing phase shift introduction methods. Furthermore, phase shift extraction is achieved using simple algorithms, each based on the number of polarizations captured by the camera. Although these algorithms may differ in their sensitivity to measurement errors \cite{malacara2007optical}, their fundamental advantages remain unchanged.

In this study, we \replaced{propose}{describe} an experimental apparatus to measure both spatial and temporal variations of gas density inside a small rectangular cavity made up of glass walls. Those walls also play the role of isothermal boundaries along which an acoustic wave is generated to analyze the thermoacoustic coupling between the oscillating gas and the rigid walls. The setup and the post-processing of experimental data are presented in Sec.\ref{sec:setup}: it comprises a polarized Mach-Zehnder interferometer equipped with a CMOS (Complementary Metal-Oxide-Semiconductor) camera providing a high spatial resolution of 1 \textmu m and a sampling frequency of $72$ Hz. The phase difference between the two beams is extracted from the fringe pattern using a four-step phase-shifting algorithm, enabling us to capture the complete 2D spatiotemporal variations of the fluid density. The experimental results are presented in Sec. \ref{sec:results}, where both the magnitude and phase of the density fluctuations are compared with the linear thermoacoustic theory for various acoustic frequencies, achieving a good agreement. As the acoustic pressure can be measured inside the cavity with a microphone, the setup also allows to determine the temperature distribution from both pressure and density measurements. The optical setup introduced in this work has the advantage of being simple to process, and it provides a 2D scan with a high spatial resolution, without the need of seeding the test cell with some smoke or fluorescent particles.

\section{\label{sec:setup}Experimental Setup}

\begin{figure}[ht]
    \centering
    \includegraphics[width=86mm]{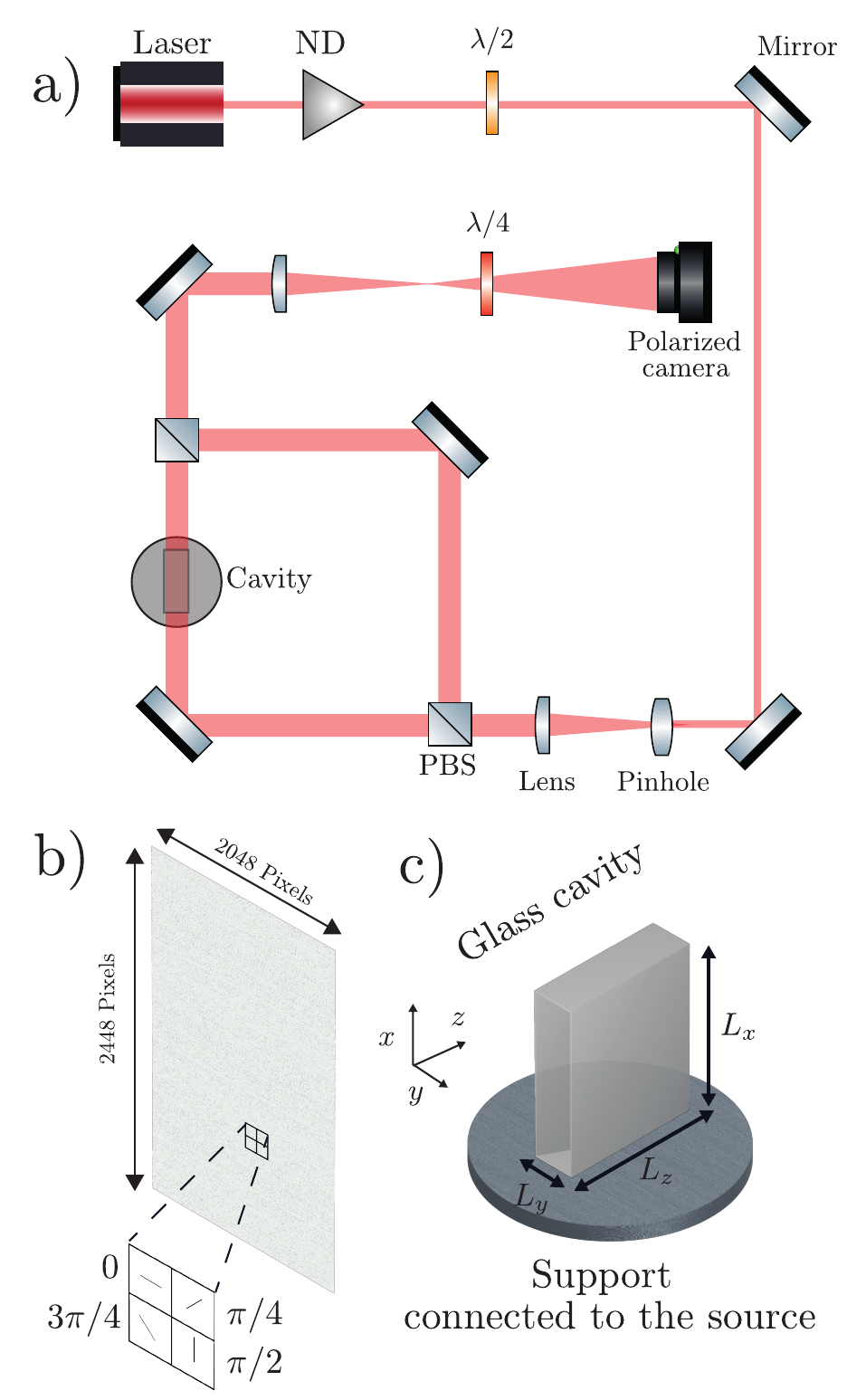}
    \caption{Polarizing Mach-Zehnder Interferometer. (a) Sketch of the experimental apparatus consisting of a Laser, Neutral Density (ND) filter, Half-Wave plates, Mirrors, Spatial Filter, Lens, Polarized Beam Splitter (PBS), Quarter-Wave Plate, and the four-polarization camera. (b) An illustration of the snapshot taken by the polarization camera, the $2048 \times 2448$ pixels image contains four different polarizations angles $[0,\pi/4,\pi/2,3\pi/4]$. (c) The test cell consisting of a thin glass cavity connected to an acoustic source, where the light beam passes through the glass along the $z$ axis.}
    \label{fig:enter-label}
\end{figure}

\subsection{Experimental Apparatus}
 
The phase-shifting Mach-Zehnder interferometer setup used to measure the density distribution is presented in Fig. 1(a). It consists of a linearly polarized Helium-Neon (He-Ne) laser (LGK7665--P18, LASOS), having a wavelength of $\lambda = 632.8$ nm, and a maximum output power of 18 mW. The emitted light passes through a neutral density filter (NDHM-50, Sigma Koki) which allows to adjust the intensity of the emitted light. Next the beam passes through a half wave-plate (10RP02-24, Newport) followed by a series of flat mirrors (10D20DM.4, Newport) used to align the light with the required optical path.

To reduce spatial noise emanating from the light source, a spatial filter is introduced, composed of an objective lens (M-20X, Newport) coupled with a pinhole aperture (900PH-25, Newport). A plano-convex lens with a focal length of $ 1000$ mm (SLB-30-1000PM, Sigma Koki) made the light collimated after passing through the spatial filter. This collimated beam is split into two beams using a polarizing beam splitter (10BC16PC.4, Newport). In the following these beams will be referred to as the test and reference beams, in which the test beam passes through the glass cavity connected to an acoustic source. 

These independent optical paths are recombined by means of another polarizing beam splitter. Here the two orthogonal polarization states are still preserved, so no interference between the two beams occurs. After recombination, the combined beam passes through a plano-convex lens (DLB-30-300PM, Sigma Koki) with a focal length of $f = 300$ mm and a quarter-wave plate (10RP04-24, Newport) set at a $\pi$/4 angle to the deflection direction of the test and reference beams. Finally, the resulting light is captured by a polarization camera (acA2440-75umPOL, Basler) with a sampling frequency up to $72$ Hz, which can be further improved by binning the images. Pixelated polarizers are placed in front of each pixel of the polarization camera, so that information corresponding to the four polarizer states (0, $\pi$/4, $\pi$/2, and 3$\pi$/4) is obtained in a single shot image of dimension $2048 \times 2448$ pixels as illustrated in Fig.1(b). After splitting the images into four interferograms of dimension $ 1024 \times 1224$ pixels with the same polarization and a spatial resolution of 1 \textmu m, the images taken are processed by a phase-shifting technique described in the next section.

The acoustic cavity, filled with air at atmospheric pressure and room temperature, has a rectangular cross-section with dimensions $L_z=50\,\text{mm}$, $L_y=10\,\text{mm}$ and a length $L_x=40\,\text{mm}$ as displayed in Fig. 1(c), and it is closed with a rigid plug. The walls are made up of glass ($1$ mm in thickness) and the optical beam is oriented parallel to the $z$-axis to characterize the thermal coupling of the oscillating gas and the rigid walls through the measured variation as a function of $x$ and $y$ of the optical path averaged along the line of sight. The cavity is placed on top of a support connected to an acoustic source (FW108N, Fostex). Herein, given the small dimension of the cavity compared to the investigated wavelengths, only the planar mode is assumed to propagate along $x$. In fact, owing to the frequency range used in the experiments described below, the oscillating pressure can even be considered as uniform in the entire cavity. In the following, the system is excited at low frequencies ($0.25$ Hz up to $10$ Hz) with amplitudes ranging from $1.5$ to $2$ kPa, and the acoustic pressure $p'$ is measured using a microphone (PD104SW-100K, JTEKT) flush-mounted at the entrance of the cavity.

\subsection{Post-Processing}
\begin{figure}
    \centering
    \includegraphics[width=86mm]{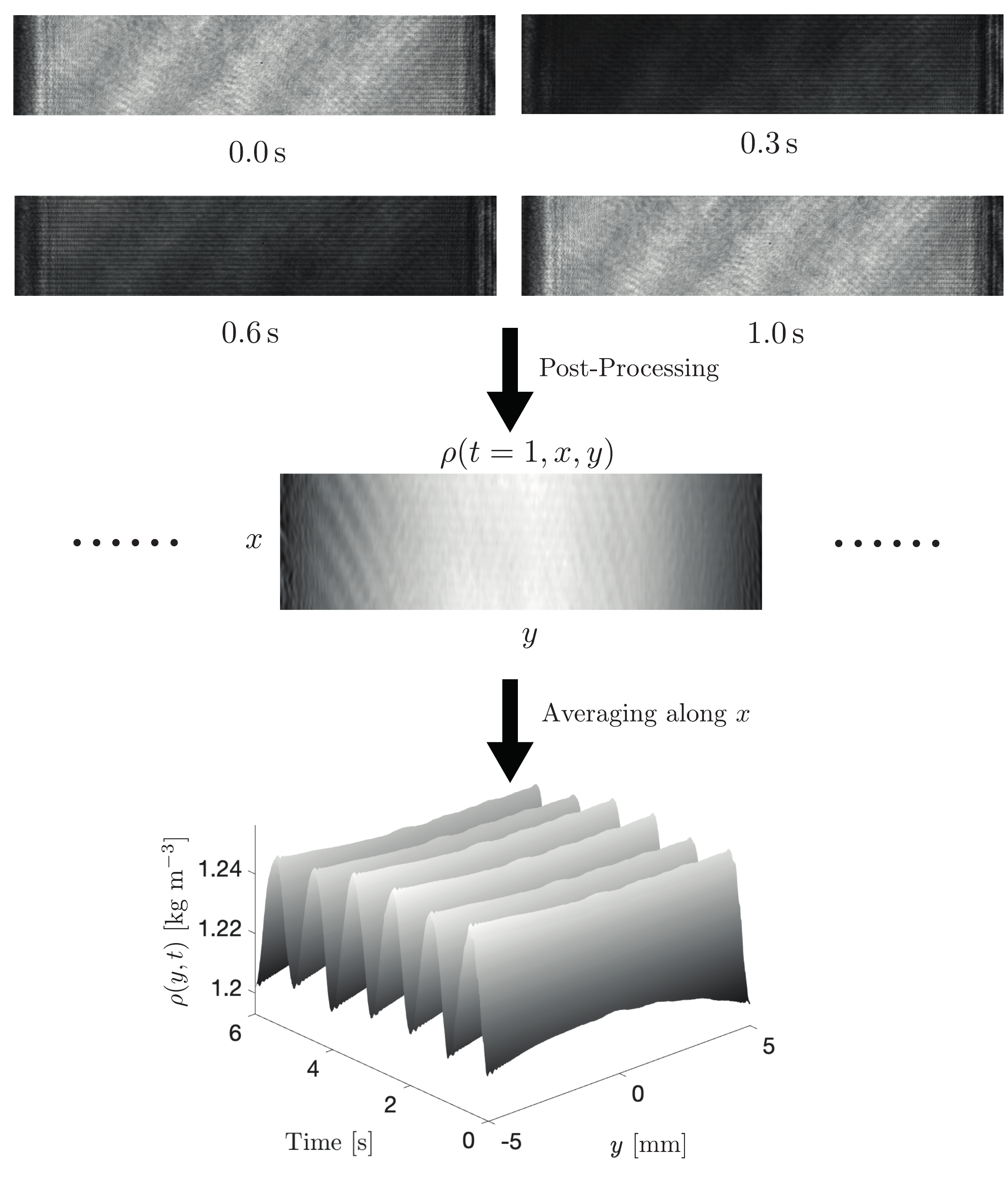}
    \caption{Illustration of the experimental results. The images taken by the polarization camera at different instant are processed using the phase-shifting algorithm, it results in a 2D spatio-temporal evolution of the density, which was eventually averaged along the $x$ axis.}
    \label{Fig2}
\end{figure}

The light amplitude $A$ passing through a polarizer with an angle $\theta$ is the superposition of the reference beam $A_\mathrm{r}$ and test beam $A_\mathrm{t}$, and it is expressed as \cite{},
$$
A(\theta, \phi)=A_\mathrm{r} \cos (\omega_0 t-\theta)+A_\mathrm{t} \cos (\omega_0 t-\theta+\phi) \text {, }
$$

\noindent
where $\omega_0$ is the angular frequency of the polarized light, and $\phi$ is the phase difference between the reference and test beams.

Herein, the polarization camera captures the spatial distribution of light intensity $I_{xy}$ at each time with four different polarizer angles $\theta_i=[0,\frac{\pi}{4},\frac{\pi}{2},\frac{3\pi}{4}]$. 

It follows that the light intensity $I_{xy}$ can be expressed as a function of the beam amplitudes, the polarization angle and the phase difference, such that \cite{torres2012development}, 

$$I_{xy}(\theta_i,\phi)=\frac{1}{2}(A_\mathrm{r}^2 +A_\mathrm{t}^2)+A_\mathrm{r} A_\mathrm{t}\cos(-\phi+2\theta_i). $$

\noindent
As the camera captures frames with four polarization, a four-step phase-shifting algorithm \cite{malacara2007optical} can be used to determine the phase difference $\phi$ at each point, and it is given by the following combination of interferograms,

\begin{equation}
    \phi(x,y)=\arctan \left[   \frac{I_{xy}(\theta_4,\phi)-I_{xy}(\theta_2,\phi)}{I_{xy}(\theta_3,\phi)-I_{xy}(\theta_1,\phi)}    \right].
\end{equation}

The resulting phase shift $\phi(x, y)$ is an arc-tangent function whose values are discontinuous and restricted to the interval $[-\pi, \pi]$. A phase unwrapping algorithm is needed to restore a continuous phase that is suitable for further analysis \cite{pijewska2019computationally,herraez2002fast}. Such algorithms have been extensively studied owing to their significant role in a number of optical devices and measurement techniques \cite{zuo2016temporal}.

The unwrapped phase shift $\phi_{uw}(x, y)$ can be used to find the refractive index $n$ of the medium \cite{hecht2012optics} inside the acoustic channel, and it is given by,

\begin{equation}
     n-n_0=\frac{\phi_{uw} \lambda}{2\pi L_z},
\end{equation}

\noindent
where $n_0$ is the refractive index of the air at rest.

With the refractive index known, the corresponding density of the fluid can be found using the Gladstone–Dale law \cite{merzkirch2012flow}, 

\begin{equation}
    \rho(t,x,y)=\rho_0\left(\frac{n-1}{n_0-1}\right),
\end{equation}

\noindent
where $\rho_0$ is the air density at ambient pressure $P_0$ and temperature $T_0$.

Since the cavity's length along $x$ is much smaller than the acoustic wavelength, the density can be assumed as uniform along the $x$-axis, such that the time variations of the density along the transverse axis $y$ can be obtained by an averaging of $M-N$ pixels along the $x$ axis.

\begin{equation}
    \rho(t,y)= \frac{1}{M-N}\sum_{j=N}^{M}  \rho(t,x_j,y)
\end{equation}

\noindent
It should be noted  that the proposed averaging along the $x$-axis is not crucial but was introduced to reduce the noise and ease the display of the results. In practice, the averaging was done on $M-N=100$ vertical pixels.

An illustration of the post-processing discussed above is proposed in Fig. 2. A portion of the raw images taken by the polarized camera are displayed at different times, and corresponds to the interferogram with angle $\theta_i=0$. Herein, the acoustic source is exciting the system at $1$ Hz. It can be observed from this raw data that the light intensities at $t=0$ s and $t=1$ s are identical, which is expected as the period of the acoustic signal is of $1$ s. Following the process described above, the refractive index inside the glass cavity is obtained by extracting the phase shift from the four interferograms using a four-step phase-shifting algorithm. The images can be successfully used to get a 2D spatio-temporal evolution of the fluid density inside the cavity. Moreover, the resulting surface can be averaged along the axis of wave propagation ($x$-axis) to get the transverse distribution of both the amplitude and the phase of density fluctuations. Note that the apparent stripes on the images can be attributed to the light reflection that can occurs on the optical elements and the glass cavity.

In the following attention is focused on the amplitude,  $|\rho'|$, and the phase, $\angle \rho'$, of the oscillating component of the density, which can be easily obtained by applying a Fourier transform of the total density. The resulting complex amplitude $\rho'=|\rho'|e^{i \angle \rho'}$ of the oscillating density can then be compared to the theory which can be derived as \cite{swift2003thermoacoustics},

\begin{equation}
    \rho'(y)= \frac{p'}{c_0^2} \{ 1+(\gamma-1)F_{\kappa}(y)\}
\end{equation}

\noindent
where $c_0$ and $\gamma$ respectively stand for the adiabatic speed of sound, the specific heat ratio of the fluid.

The $y$-dependent function  $F_{\kappa}$ takes into account the effect of the thermal boundary layers and depends on the geometry, on the angular frequency of the acoustic wave $\omega$,  and on the thermal diffusivity of the fluid $\kappa$. For the case of two parallel plates separated by a length $L_y$, the thermal function is given by,

\begin{equation}
    F_{\kappa} =  \frac{\cosh[(1+i)y/\delta_\kappa]}{\cosh [(1+i)L_y/2\delta_\kappa]},
\end{equation}

\noindent
where the thermal boundary layer thickness is given by $\delta_{\kappa}=\sqrt{\frac{2\kappa}{\omega}}$.

Furthermore, since the acoustic density $\rho'$ caused by the sound wave is measured, and since the acoustic pressure $p'$ is also measured using a microphone, then the spatial distribution of the temperature oscillations $T'$ can be obtained from the ideal gas law

\begin{equation}
    T'(y)=T_0\Bigl\{\frac{p'}{P_0}-\frac{\rho'(y)}{\rho_0}\Bigl\},
\end{equation}
 and compared with theory\cite{swift2003thermoacoustics}.

\section{\label{sec:results}Experimental Results}

\begin{figure*}
    \centering
    \includegraphics[width=176mm]{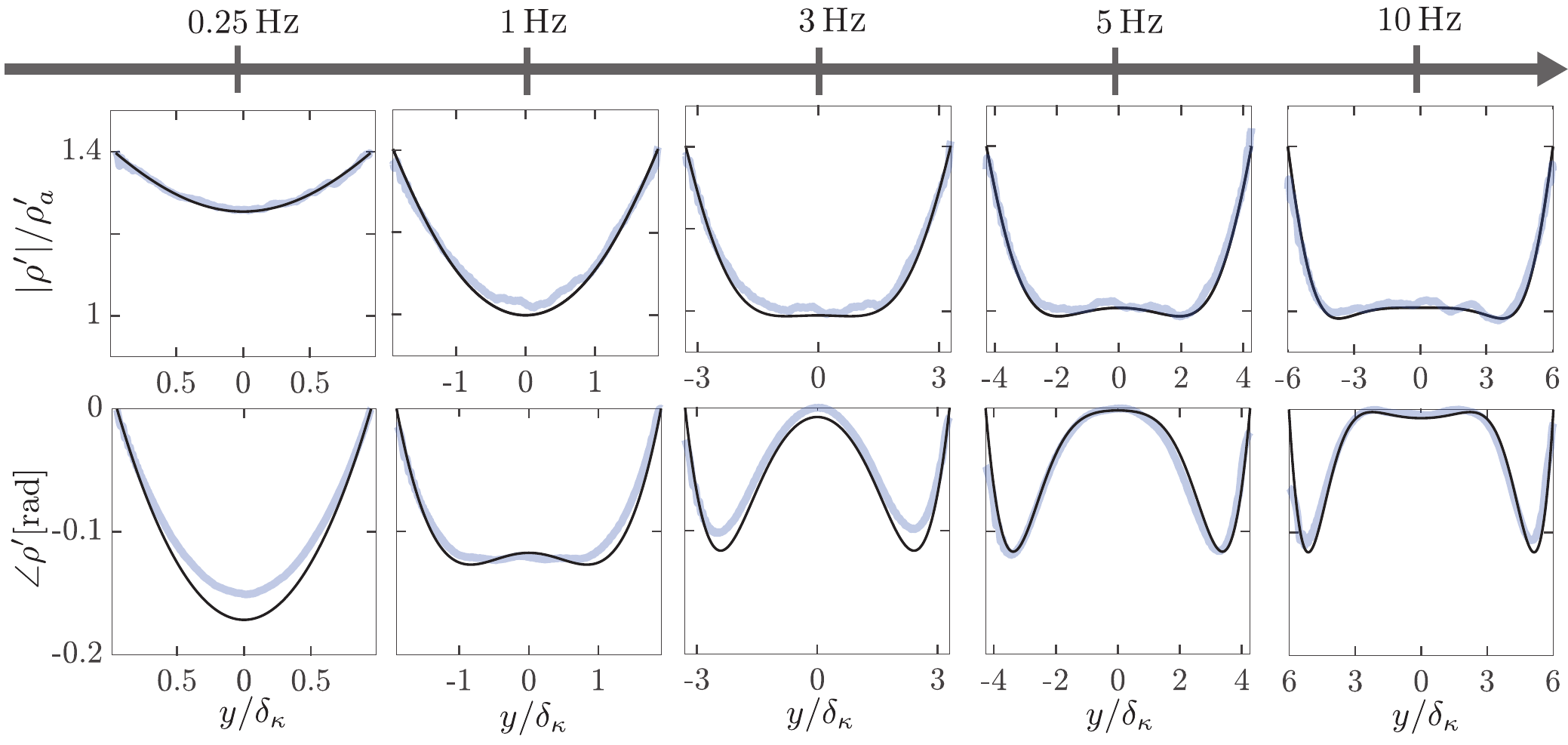}
    \caption{Measured density oscillations. The normalized amplitude and the phase of the time-varying density relative to the pressure is displayed as a function of the normalized position $y/\delta_{\kappa}$ for frequencies ranging from $0.25$ Hz to $10$ Hz. The blue and black lines represent the experimental results using the proposed apparatus and the theoretical result, respectively. }
    \label{Fig3}
\end{figure*}

In this section, the experimental results obtained using the phase-shifting interferometer are discussed and compared with the theoretical results derived from the thermoacoustic theory.

Figure 3 shows the amplitude and phase of the density distribution measured along the $y$ axis using the interferometer, which are also compared with the theoretical results. The experiments were performed at various frequencies ranging from $0.25$ Hz to $10$ Hz and are represented by light blue lines, while theoretical results are shown in black. Furthermore, as the underlying phenomenon is assumed linear such that the spatial distribution does not depend on the amplitude of the forcing, a normalization of the density amplitude by the adiabatic density $\rho'_a=\frac{p'}{c_0^2}$  is applied to compare more easily the data obtained at different frequencies (the amplitude of the acoustic pressure being slightly different from one set of measurements to another one at another frequency). Moreover, the $y$-axis is also scaled by $\delta_\kappa$ to better illustrate the impact of the boundary layer on the spatial distribution of density oscillations.

The range of frequencies at which the cavity is excited was chosen based on some experimental constraints regarding the lower frequency limit of the loudspeaker and the sampling rate of the camera. As a result, the distance $L_y$ between the two rigid walls was chosen such that the thermal coupling between the oscillating gas and the walls ranges from a quasi-isothermal process, namely $L_y \approx 1.69 \delta_\kappa$ at 0.25 Hz, to a quasi-adiabatic process, namely $L_y \approx 9 \delta_\kappa$ at $10$ Hz. Therefore, the investigated distributions of density fluctuations are typical of those found in the heat exchangers or in the stack of standing wave thermoacoustic heat engines, for which the distance between two plates is of a few thermal boundary layer thicknesses. The theory (black lines) predicts that during the transition from a quasi-isothermal to a quasi-adiabatic motion, the transverse distribution of density oscillations passes from a quasi-parabolic profile (at $0.25$ Hz) to a quasi-planar profile (at $10$ Hz) where the maximum of density is not located on the median axis. As shown in Fig. \ref{Fig3}, those profiles are almost perfectly reproduced in the experiments, while the $y$-dependency of the phase of density fluctuations is also in very good agreement with theory. At higher frequencies and near the walls, we can observe a small lack of precision, which may be caused by a slight misalignment of the cavity with respect to the light beam, causing an unwanted refraction of the light passing through a gradient of density.
\begin{figure}
    \centering
    \includegraphics[width=86mm]{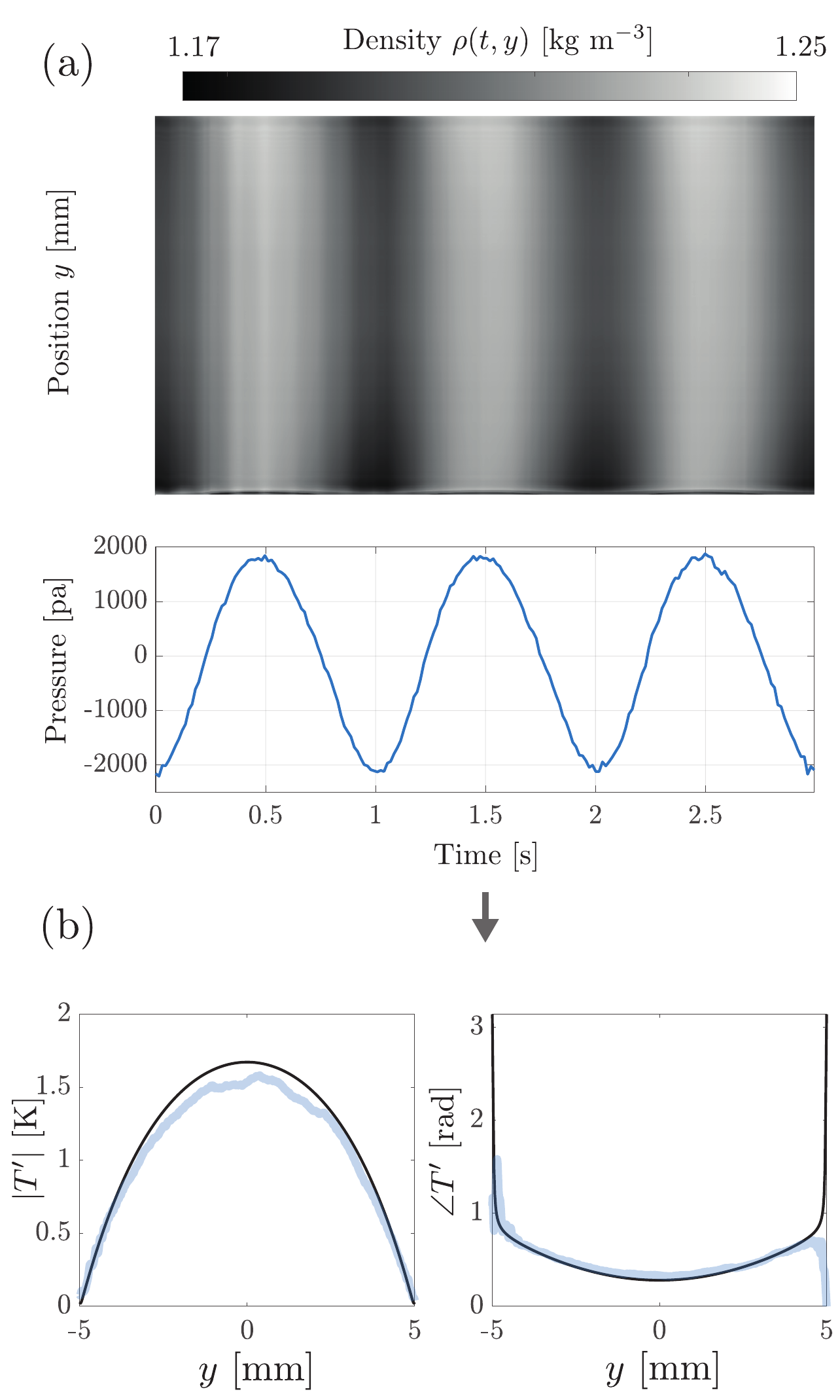}
    \caption{Estimation of the temperature fluctuation $T'$ at $f=1$ Hz. (a) The synchronous measurements of the spatio-
temporal density and time-varying pressure. (b) The corresponding amplitude and phase  of temperature oscillations as a function of the position $y$. Blue and black lines represents respectively the experimental and theoretical results.  }
    \label{Temp}
\end{figure}

Figure \ref{Temp}(a) displays the synchronous temporal evolution of the density and the pressure for the results displayed in Fig. 3 with $f=1$ Hz. As discussed previously, by combining the two measurements, the ideal gas law (See Eq.7) can be used to retrieve the spatial distribution of the oscillating temperature. Figure \ref{Temp}(b) shows the amplitude and phase of temperature oscillations $T'$ as a function of the position along the $y$ axis, where black and blue lines represent the theory and experimental results, respectively. The results show that at the center of the cavity the temperature oscillation exceeds $1.5$ K in amplitude, while it vanishes next to the wall, as expected. Furthermore, the phase of the temperature fluctuation is well described at the center of the cavity. Yet, it is difficult to accurately estimate it in the immediate vicinity of the wall owing to the unavoidable diffraction and errors in density measurements in this region. Nevertheless, the amplitude and phase results shown indicate that theory and experiment are in good agreement. Although similar results were already partially obtained in previous studies (for temperature fluctuations) using either a cold wire sensor \cite{huelsz1998temperature}, a PLIF technique \cite{shi2010application}, or with holographic interferometry \cite{wetzel2000experimental}, the experimental results presented in Figs. \ref{Fig3} and \ref{Temp}(b) show that the phase-shifting interferometry is an efficient way to measure accurately the density fluctuations, even next to solid boundaries where heat exchanges are responsible for significant gradients of density. As this technique is also a full-field measurement which does not need seeding particles, it could be used advantageously for the investigation of poorly described heat exchange phenomena like the ones involved at the connection between the stack and the heat-exchangers of a thermoacoustic heat engine.


\section{Conclusion}

In this work, a phase-shifting Mach-Zehnder interferometer was developed to measure the density fluctuations generated by an acoustic wave between two parallel plates. The profiles of the oscillating density were successfully measured by extracting the phase difference between the reference and test beams using a four-step phase-shifting algorithm. Additionally, by combining density and pressure measurements, we successfully retrieved the temperature fluctuations using the ideal gas law. Hence, the experimental setup allows us to get the temperature, the density, and the pressure inside the cavity.

The experimental results obtained through this approach show a very good potential and could be extended to study more sophisticated phenomena, like for instance the entrance effects and associated nonlinear heat transport involved at the connection between the stack and the heat exchangers in thermoacoustic heat pumps or engines \cite{berson2011nonlinear}. Moreover, these results can be improved further by incorporating filters to reduce the different types of noises \cite{servin2009noise}, by refining the alignment of optical elements, or by using a higher order phase-shifting algorithm \cite{zhang1999error,hack2011invited}. Also, for the analysis of steady-state oscillations, the limits in terms of sampling rate (a few tens of Hertz in this case) due to the Shannon criterion, could be pushed back by using a phase-locking prodecure. Finally, it would be worth considering combining this measurement technique with another one giving access to the velocity (e.g. PIV) which would enable to get a direct measurement of the convective transport of heat inherent to the thermoacoustic effect.

\section*{Acknowledgments}
M.A received financial support from the Institut d'Acoustique Graduate School of Le Mans (IA-GS) for his mobility to Japan.
This work was supported by JST FOREST Program, Grant Number JPMJFR223P; JSPS Grant-in-Aid for Exploratory Research, Grant Number 23K17838.

\section*{Author Declarations}
\subsection*{Conflict of Interest}
The authors have no conflicts of interest to declare
that are relevant to the content of this article.

\section*{Data Availability}
The data that support the findings of this study are
available from the corresponding author upon reasonable
request.

\section*{Copyright}

The following article has been accepted by the Journal of Acoustical Society of America (JASA). After it is published, it will be found at https://pubs.aip.org/asa/jasa.

\bibliography{sampbib}
    
\end{document}